\documentclass[a4paper,11pt]{article}

\usepackage{graphicx,epsfig,pstricks,fancybox,fancyhdr,epic,rotating,color}
\usepackage{amsfonts}

\begin{document}

\setcounter{page}{0}
\thispagestyle{empty}
\begin{titlepage}

\vspace*{-1cm}
\hfill 
19. April, 2008 
   
\vfill

\begin{center}
  {\large \bf
 On the class of chiral symmetry representations \\
 with scalar and pseudoscalar
 fields
      } 
\vfill
\vspace*{0.3cm} 

    Peter Minkowski  \\
    Institute for Theoretical Physics \\
    University of Bern \\
    CH - 3012 Bern, Switzerland
    \\
    E-mail: mink@itp.unibe.ch
   \vspace*{0.3cm} \\  

\end{center}

\vfill

\begin{abstract}
\noindent
In the following few pages an account is given of a theme , which
I began in 1966 and continued to the present.
\end{abstract}

\vfill
\end{titlepage}

\section{$\Sigma \ = \ \frac{1}{\sqrt{2}} 
\ \left ( \ \sigma \ - \ i \ \pi \ \right )$ scalar - pseudoscalar fields and
the class of their chiral symmetry representations
}
\label{two}

\noindent
Lets denote by $t \ , \ s \ , \ n \ \cdots$ quark flavor indices with

\vspace*{-0.1cm}
\begin{equation}
\label{cheq:1}
\begin{array}{l}
t \ , \ s \ , \ n \ \cdots \ = \ 1 \ , \ \cdots \ , \ N \ \equiv \ N_{\ fl}
\end{array}
\end{equation}

\noindent
and by $\overline{\lambda}^{\ a}$ the $N^{\ 2}$ hermitian 
$U_{\ N}$ matrices with the normalization

\vspace*{-0.1cm}
\begin{equation}
\label{cheq:2}
\begin{array}{l}
\overline{\lambda}^{\ a} \ = 
\ \left ( \ \overline{\lambda}^{\ a} \ \right )_{\ ts}
\hspace*{0.2cm} ; \hspace*{0.2cm}
tr \ \overline{\lambda}^{\ a} \ \overline{\lambda}^{\ b} \ = \ \delta_{\ a b}
\vspace*{0.2cm} \\
a \ = \ 0 \ , \ 1 \ , \ \cdots \ , \ N^{\ 2} \ - \ 1
\hspace*{0.2cm} ; \hspace*{0.2cm} 
\overline{\lambda}^{\ 0} \ = \ \sqrt{N}^{\ - \ 1/2} 
\ \left ( \ \P \ \right )_{\ N \times N}
\vspace*{0.2cm} \\
tr \ \overline{\lambda}^{\ a} \ = \ 0 \hspace*{0.2cm}  \mbox{for} 
\hspace*{0.2cm} a \ > \ 0
\hspace*{0.2cm} ; \hspace*{0.2cm} 
\left . \lambda^{\ a} \ \right |_{\ conv.} \ = 
\ \sqrt{2} \hspace*{0.2cm} \overline{\lambda}^{\ a} 
\end{array}
\end{equation}

\noindent
In order to maintain clear quark field association we choose 
the convention {\it and restriction} projecting out color and spin degrees
of freedom from the complete set of $\overline{q} \ q$ bilinears

\vspace*{-0.1cm}
\begin{equation}
\label{cheq:3}
\begin{array}{l}
\Sigma_{\ s \ \dot{t}} \ \sim \ \overline{q}_{\ \dot{t}}^{\ \dot{c}}
\ \frac{1}{2} \ \left ( \ 1 \ + \ \gamma_{\ 5 \ R} \ \right ) \ q_{\ s}^{\ c}
\vspace*{0.2cm} \\
\gamma_{\ 5 \ R} \ = \ \frac{1}{i} \ \gamma_{\ 0} \ \gamma_{\ 1} 
\ \gamma_{\ 2} \ \gamma_{\ 3} 
\hspace*{0.3cm} ; \hspace*{0.3cm}
c , \dot{c} \ = \ 1,2,3 \hspace*{0.2cm} \mbox{color}
\end{array}
\end{equation}

\noindent
The {\it logical} structure of $\Sigma$ - variables is different, when used 
to {\it derive} the dynamics of quarks, i.e. QCD, or before this,
when used in their own right as by M. Gell-Mann and M. L\'{e}vy \cite{LevyGM} , 
or else associating chiral symmetry with superconductivity 
as by Y. Nambu and G. Jona-Lasinio \cite{Nambu}.
\vspace*{0.1cm}

\noindent
Here the chiral \hspace*{0.1cm} $UN_{\ fl \ R} \ \times \ UN_{\ fl \ L}$
\hspace*{0.1cm} transformations correspond to

\vspace*{-0.1cm}
\begin{equation}
\label{cheq:4}
\begin{array}{l}
\begin{array}{c}
UN_{\ fl \ R} \ : \ \frac{1}{2} 
\ \left ( \ 1 \ + \ \gamma_{\ 5 \ R} \ \right ) \ q_{\ s}^{\ c}
\rightarrow \ V_{\ s s'} \hspace*{0.2cm} \frac{1}{2} 
\ \left ( \ 1 \ + \ \gamma_{\ 5 \ R} \ \right ) 
\ q_{\ s'}^{\ c} 
\vspace*{0.2cm} \\
UN_{\ fl \ L} \ :
\ \frac{1}{2}
\ \left ( \ 1 \ - \ \gamma_{\ 5 \ R} \ \right ) \ q_{\ s}^{\ c}
\ \rightarrow \ W_{\ s s'} \ \frac{1}{2} 
\ \left ( \ 1 \ - \ \gamma_{\ 5 \ R} \ \right )
\ q_{\ s'}^{\ c}
\vspace*{0.2cm} \\
\updownarrow
\vspace*{0.2cm} \\
\Sigma \ \rightarrow \ V \ \Sigma \ W^{\ -1}
\end{array}
\end{array}
\end{equation}

\noindent
The construction in eq. \ref{cheq:4} can be interpreted 
as group-complexification , discussed below.
The $\Sigma$-variables arise as 
classical field configurations ,
Legendre transforms of the QCD generating functional driven by general 
x-dependent complex {\it color neutral} mass terms. 

\newpage

\noindent
The latter represent external sources with 
\hspace*{0.1cm} $UN_{\ fl \ R} \ \times \ UN_{\ fl \ L}$
\hspace*{0.1cm} substitutions aligned with the $\Sigma$ - variables

\vspace*{-0.1cm}
\begin{equation}
\label{cheq:5}
\begin{array}{l}
\begin{array}{lll}
- \ {\cal{L}}_{\ m} & = & m_{\ \dot{t} s} \ ( \ x \ )
\ \left \lbrace \ \overline{q}_{\ s}^{\ \dot{c}} 
\ \frac{1}{2} \ \left ( \ 1 \ - \ \gamma_{\ 5 \ R} \ \right ) \ q_{\ t}^{\ c}
\ \right \rbrace \ + \ h.c.
\vspace*{0.2cm} \\
& \propto & tr \left ( \ m \ \Sigma^{\ \dagger} \ + \ \Sigma \ m^{\ \dagger}
\ \right )
\end{array}
\vspace*{0.2cm} \\
m \ \rightarrow \ V \ m \ W^{\ -1}
\ \longleftrightarrow
\ \Sigma \ \rightarrow \ V \ \Sigma \ W^{\ -1}
\end{array}
\end{equation}

\noindent
The so defined (classical) target space variables 
\footnote{\hspace*{0.1cm}
The notion of target-space is used as defined in modern context of
string theories .} form 

\begin{center}
-- {\it upon the exclusion of values for which} $Det \ \Sigma \ = \ 0$ --
\end{center}
the group

\vspace*{-0.1cm}
\begin{equation}
\label{cheq:6}
\begin{array}{l}
GL \ ( \ N \ , \ C \ ) \hspace*{0.2cm} = \hspace*{0.2cm}
\left \lbrace \left . \ \Sigma \ \right | \ Det \ \Sigma \ \neq \ 0 
\ \right \rbrace
\end{array}
\end{equation}

\noindent
the general linear group
over the complex numbers in N dimensional
target-space .

\noindent
We proceed to define the hermitian chiral currents generating \\
$UN_{\ fl \ R} \ \times \ UN_{\ fl \ L}$ ( global )
pertaining to $\Sigma$

\vspace*{-0.1cm}
\begin{equation}
\label{cheq:7}
\begin{array}{l}
\begin{array}{lll}
j^{\ a}_{\ \mu \ R} \ = \ tr \ \Sigma^{\ \dagger}
\ \left ( \ \frac{1}{2} \ \lambda^{\ a} \ i 
\ \stackrel{\rightleftharpoons}{\partial}_{\ \mu} \ \right )
\ \Sigma
& \sim & \ \overline{q} \ \gamma_{\ \mu} \ \frac{1}{2} \ \lambda^{\ a}
P_{\ R} \ q
\vspace*{0.2cm} \\
j^{\ a}_{\ \mu \ L} \ = \ tr \ \Sigma^{\ \dagger} 
\ i \ \stackrel{\rightleftharpoons}{\partial}_{\ \mu} \ \Sigma
\ \left ( \ - \ \frac{1}{2} \ \lambda^{\ a} \ \right )
& \sim & \ \overline{q} \ \gamma_{\ \mu} \ \frac{1}{2} \ \lambda^{\ a}
P_{\ L} \ q
\end{array}
\vspace*{0.2cm} \\
A \ \stackrel{\rightleftharpoons}{\partial}_{\ \mu} \ B
\ = \ A \ \partial_{\ \mu} \ B \ - \ ( \ \partial_{\ \mu} \ A \ ) \ B
\hspace*{0.2cm} ; \hspace*{0.2cm}
P_{\ R \ (L)} \ = \ \frac{1}{2} \ \left ( \ 1 \ \pm \ \gamma_{\ 5 \ R} 
\ \right )
\end{array}
\end{equation}

\noindent
We avoid here to couple external sources to all other 
$\overline{q} \ q$ bilinears except the scalar - pseudoscalar ones as specified
in eq. \ref{cheq:5} for two reasons

\begin{description}
\item 1) -- to retain a minimum set of external sources capable to reproduce
spontaneous {\it chiral} symmetry breaking alone as a restricted but
fully dynamical spontaneous phenomenon.

\item 2) -- in order to avoid a nonabelian anomaly structure .
The latter would force either the consideration of leptons in addition to
quarks , or the inclusion of nonabelian Wess-Zumino terms obtained
from connections formed from the $\Sigma$ fields \cite{Wesszwand} .

\end{description}

\newpage

\noindent
For completeness we display the equal time current algebra relations inherited
from  $\overline{q} \ q$ 

\vspace*{-0.1cm}
\begin{equation}
\label{cheq:8}
\begin{array}{l}
\left \lbrack \ j_{\ 0 \ R}^{\ a} \ ( \ t \ , \ \vec{x} \ ) \ ,
\ j_{\ 0 \ R}^{\ b} \ ( \ t \ , \ \vec{y} \ ) \ \right \rbrack
\ = \ i \ f_{\ a b n} \ j_{\ 0 \ R}^{\ n} \ ( \ t \ , \ \vec{x} \ )
\ \delta^{\ 3} \ ( \ \vec{x} \ - \ \vec{y} \ )
\vspace*{0.2cm} \\
\left \lbrack \ j_{\ 0 \ L}^{\ a} \ ( \ t \ , \ \vec{x} \ ) \ ,
\ j_{\ 0 \ L}^{\ b} \ ( \ t \ , \ \vec{y} \ ) \ \right \rbrack
\ = \ i \ f_{\ a b n} \ j_{\ 0 \ L}^{\ n} \ ( \ t \ , \ \vec{x} \ )
\ \delta^{\ 3} \ ( \ \vec{x} \ - \ \vec{y} \ )
\vspace*{0.2cm} \\
\left \lbrack \ j_{\ 0 \ R}^{\ a} \ ( \ t \ , \ \vec{x} \ ) \ ,
\ j_{\ 0 \ L}^{\ b} \ ( \ t \ , \ \vec{y} \ ) \ \right \rbrack
\ = \ 0
\vspace*{0.2cm} \\
\left \lbrack \ \frac{1}{2} \ \lambda^{\ a} \ ,
\ \frac{1}{2} \ \lambda^{\ b} \ \right \rbrack \ = \ i \ f_{\ a b n}
\ \frac{1}{2} \ \lambda^{\ n}
\end{array}
\end{equation}

\noindent
The $GL \ ( \ N \ , \ C \ )$ group structure 
defined in eq. \ref{cheq:6} enables 
bilateral multiplication of the $\Sigma \ , \ Det \ \Sigma \ \neq \ 0$
elements , of which the left- and right-chiral currents defined in 
eq. \ref{cheq:7} are {\it naturally} associated with the Lie-algebra
of $UN_{\ fl \ R} \ \times \ UN_{\ fl \ L}$ through the exponential mapping
with subgroups of 
$GL \ ( \ N \ , \ C \ )_{\ R} \ \times \ GL \ ( \ N \ , \ C \ )_{\ L}$ .
These (sub)groups act by multiplication of the base-group-manifold
by respective multiplication from the 
$\mbox{left} \ \leftrightarrow \ G_{\ R}$ and from the 
$\mbox{right} \ \leftrightarrow \ G_{\ L}$ . The reverse association \\
-- here -- is {\it accidental}

\vspace*{-0.1cm}
\begin{equation}
\label{cheq:9}
\begin{array}{l}
GL \ ( \ N \ , \ C \ )_{\ R \ (L)} \ \rightarrow \ G_{\ R \ (L)} \ = \ G
\vspace*{0.2cm} \\
\Sigma \ \in \ G 
\hspace*{0.2cm} ; \hspace*{0.2cm}
g \ \in \ G_{\ R} 
\hspace*{0.2cm} ; \hspace*{0.2cm} 
h \ \in \ G_{\ L} \ : 
\vspace*{0.2cm} \\
\begin{array}{rll}
G_{\ R} \ \bullet \ G & \leftrightarrow & \Sigma \ \rightarrow \ g \ \Sigma
\vspace*{0.2cm} \\
G_{\ L} \ \bullet \ G & \leftrightarrow & \Sigma \ \rightarrow 
\ \Sigma \ h^{\ -1}
\vspace*{0.2cm} \\
G_{\ R} \ \otimes \ G_{\ L} \ \bullet \ G
& \leftrightarrow & \Sigma \ \rightarrow \ g \ \Sigma \ h^{\ -1}
\end{array}
\vspace*{0.2cm} \\
\Sigma \ = \ \Sigma \ ( \ x \ ) 
\hspace*{0.2cm} ; \hspace*{0.2cm} 
g \ , \ h \ : \ \mbox{x-independent or 'rigid'}
\end{array}
\end{equation}

\begin{center}
{\bf The exponential mapping and compactification(s) of 
$G \ ( \ \Sigma \ )$}
\end{center}

\noindent
The condition $Det \ \Sigma \ \neq \ 0$ in the restriction to 
$GL \ ( N \ , C \ )$ ( eq. \ref{cheq:6} ) is very special
and surprising in conjunction with the field variable definition.
\vspace*{0.2cm}

\noindent
In fact such a condition is completely untenable and shall be
discussed below. This was a stumbling block for a while .

\newpage

\noindent
This condition is equivalent to the relation with the Lie algebra \\
of $GL \ ( \ N \ , \ C \ )$ through the exponential mapping and
its inverse ( $\log$ )

\vspace*{-0.1cm}
\begin{equation}
\label{expeq:1}
\begin{array}{l}
\Sigma \ = \ \exp \ b
\hspace*{0.2cm} ; \hspace*{0.2cm}
b \ = \ b^{\ a} \ \frac{1}{2} \ \lambda^{\ a}
\hspace*{0.2cm} ; \hspace*{0.2cm}
\frac{1}{2} \ \lambda^{\ 0} \ = \ ( \ 2 \ N \ )^{\ -1/2} 
\ \left ( \ \P \ \right )_{\ N \times N}
\vspace*{0.2cm} \\
Det \ \Sigma \ = \ \exp \ ( \ tr \ b \ ) \ = \ \exp \ \beta
\hspace*{0.2cm} ; \hspace*{0.2cm}
\beta \ = \ \sqrt{\ \frac{2}{N} \ } \ b^{\ 0} 
\vspace*{0.2cm} \\
Det \ \Sigma \ = 0 \ \leftrightarrow \ \Re \ \beta \ = \ - \ \infty
\hspace*{0.2cm} ; \hspace*{0.2cm}
\beta \ \sim \ \beta \ + \ 2 \ \pi \ i \ \nu
\hspace*{0.2cm} ; \hspace*{0.2cm}
\nu \ \in \ \mathbb{Z}
\end{array}
\end{equation}

\noindent
Of course eliminating 
-- from general {\it dynamical} $\Sigma$-variables --
the subset with $Det \ \Sigma \ = \ 0$ affects only the non-solvable  
( and non-semi-simple \footnote{\hspace*{0.1cm} The words testify to the fight
for definite mathematical notions .} )
part of the associated group, whence the former are
interpreted as a manifold, 
which simply is {\it not} a group .
It may thus appear that the restriction
in order to enforce a group structure is characterized by the
notion of 'group-Plague', infecting the general structure at hand .

\noindent
This said we continue to treat $\Sigma$-variables as if they were identifiable
with $GL \ ( \ N \ , \ C \ )$ .

\noindent
The next reductive step is to consider the solvable ( simple ) subgroup

\vspace*{-0.1cm}
\begin{equation}
\label{expeq:2}
\begin{array}{l}
SL \ ( \ N \ , \ C \ ) \ \subset \ GL \ ( \ N \ , \ C \ ) \ \subset
\ \left \lbrace \ \Sigma \ \right \rbrace 
\vspace*{0.2cm} \\
SL \ ( \ N \ , \ C \ ) \ = \ \left \lbrace \ \left . \widehat{\Sigma} 
\ \right | \ Det \ \widehat{\Sigma} \ = \ 1 \ \right \rbrace
\vspace*{0.2cm} \\
\widehat{\Sigma} \ \sim \ \Sigma \ / 
\ \left ( \ Det \ \Sigma \ \right )^{\ 1/N}
\hspace*{0.2cm} ; \hspace*{0.2cm}
\mbox{allowing {\it all} N roots}
\end{array}
\end{equation}

\noindent
The advantage of the above reduction to $SL \ ( \ N \ , \ C \ )$
is that it allows the exponential mapping to an irreducible ( simple )
Lie-algebra , \\
refining eq. \ref{expeq:1} 

\vspace*{-0.1cm}
\begin{equation}
\label{expeq:3}
\begin{array}{l}
\widehat{\Sigma} \ = \ \exp \ \widehat{b}
\hspace*{0.2cm} ; \hspace*{0.2cm}
\widehat{b} \ = \ \widehat{b}^{\ a} \ \frac{1}{2} \ \lambda^{\ a}
\hspace*{0.2cm} ; \hspace*{0.2cm}
a \ = \ 1 \ , \ 2 \ , \ \cdots \ , \ N^{\ 2} \ - \ 1
\vspace*{0.2cm} \\
\widehat{b}^{\ 0} \ = \ 0 
\hspace*{0.2cm} ; \hspace*{0.2cm}
tr \ \lambda^{\ a} \ = \ 0
\end{array}
\end{equation}

\noindent
i.e. eliminating the unit matrix \hspace*{0.1cm} 
$\propto \ \lambda^{0}$ \hspace*{0.1cm} from the latter .

\newpage

\begin{center}
{\bf 1.1 Relaxing the condition 
$Det \ \Sigma \ \neq \ 0$ and the unique association 
\vspace*{0.2cm} \\
$\Sigma \hspace*{0.2cm} 
\begin{array}[t]{c} 
{\scriptstyle
\longrightarrow}
\vspace*{-0.2cm} \\
{\scriptstyle
Det \ \Sigma \ \neq \ 0}
\end{array}
\hspace*{0.2cm}  GL \ ( \ N \ , C \ )$}
\end{center}

\noindent
We transform $\Sigma_{\ s \dot{t}}$ as defined or better associated
in eq. \ref{cheq:3} by means of the $N^{\ 2}$ {\it hermitian} matrices
$\overline{\lambda}^{\ a}$ in eq. \ref{cheq:2} .

\vspace*{-0.1cm}
\begin{equation}
\label{cneq:1}
\begin{array}{l}
\Sigma_{\ s \dot{t}} \ = \ \Sigma^{\ a} 
\ \left ( \ \overline{\lambda}^{\ a} \ \right )_{\ s \dot{t}}
\vspace*{0.2cm} \\
\Sigma^{\ a} \ = \ tr \ \overline{\lambda}^{\ a} \ \Sigma
\hspace*{0.2cm} ; \hspace*{0.2cm}
a \ = \ 0 \ , \ 1 \ , \ \cdots \ , \ N^{\ 2} \ - \ 1
\end{array}
\end{equation}

\noindent
The complex ( field valued ) quantities $\Sigma^{\ a}$ are components
of a complex $N^{\ 2}$-dimensional space $C_{\ N^{\ 2}}$ and in one to one
correspondence with the matrix elements $\Sigma_{\ s \dot{t}}$ 

\vspace*{-0.1cm}
\begin{equation}
\label{cneq:2}
\begin{array}{l}
C_{\ N^{\ 2}} \ = \ \left \lbrace \ \left ( \ \Sigma^{\ 0} \ , 
\ \Sigma^{\ 1} \ , \ \cdots \ \Sigma^{\ N^{\ 2} -1} \ \right ) \ \right \rbrace
\end{array}
\end{equation}

\noindent
This serves to become aware of the {\it second} algebraic 
relation ( $\oplus$ ) , beyond ( $\otimes$ ) ,
i.e. to add matrices and not to just multiply them .
\footnote{\hspace*{0.1cm} Elements of a $N \ \times \ N$-matrix can 
equivalently be arranged along a line .}
\vspace*{0.1cm}

\noindent
The $\oplus$ operation is {\it also} encountered upon 'shifting'
general (pseudo)scalar fields relative to a spontaneous vacuum expected
value . This is relevant {\it here} 
for spontaneous breaking of chiral symmetry .  
\vspace*{0.1cm}

\noindent
It arises independently for the $SU2_{\ L}$-doublet scalar (Higgs) fields .
\vspace*{0.1cm}

\noindent
Hence the idea that the combination of $\oplus$ and  $\otimes$ -- which form
the full motion group ( of matrices ) -- are related to 'fields'
( 'K\"{o}rper' in german ). Thus we are led to consider quaternion-
and octonion-algebras in the next sections .



\begin{center}
{\bf 1.2 Octonions ( or Cayleigh numbers ) as pairs of quaternions}
\end{center}

Let 

\vspace*{-0.3cm}
\begin{equation}
\label{octeq:1}
\begin{array}{l}
q \ = q^{\ 0} \ i_{\ 0} \ + \ q^{\ a} \ i_{\ a} 
\hspace*{0.2cm} ; \hspace*{0.2cm} 
a \ = \ 1,2,3 
\hspace*{0.2cm} ; \hspace*{0.2cm} \left ( \ q^{\ 0} \ , \ \vec{q} \ \right )
\ \in \ R_{\ 4}
\vspace*{0.2cm} \\
i_{\ 0} \ = \ \P 
\hspace*{0.2cm} ; \hspace*{0.2cm}
i_{\ a} \ i_{\ b} \ = \ - \ \delta_{\ ab} \ i_{\ 0}
\ + \ \varepsilon_{\ a b n} \ i_{\ n}
\hspace*{0.4cm} | \ \mbox{for} \hspace*{0.2cm}
a \ , \ b \ , \ n \ = \ 1,2,3
\vspace*{0.2cm} \\
\overline{q} \ = \ q^{\ 0} \ i_{\ 0} \ - \ q^{\ a} \ i_{\ a}
\end{array}
\end{equation}

\noindent
denote a quaternion over the real numbers .

\noindent
Then a single octonion is represented ( modulo external automorphisms
\footnote{\hspace*{0.1cm} These automorphisms form the exceptional group 
$G_{\ 2}$ .})
by a pair of quaternions $( \ p \ , \ q \ )$  with the 
nonassociative multiplication rule

\vspace*{-0.3cm}
\begin{equation}
\label{octeq:2}
\begin{array}{l}
o \ = \ \left ( \ p \ , \ q \ \right ) \ = \ p^{\ 0} \ j_{\ 0}
\ + \ p^{\ a} \ j_{\ a} \ + \ q^{\ 0} \ j_{\ 4} \ + \ q^{\ a} \ j_{\ 4 \ + \ a}
\vspace*{0.2cm} \\
o^{\ \alpha} \ = \ \left ( \ p^{\ \alpha} \ , \ q^{\ \alpha} \ \right )
\hspace*{0.2cm} ; \hspace*{0.2cm}
\alpha \ = \ 1,2,\cdots
\vspace*{0.2cm} \\
o^{\ 1} \ \odot \ o^{\ 2} \ =
\left ( \ p^{\ 1} \ p^{\ 2} \ - \ \overline{q}^{\ 2} \ q^{\ 1}
\ , \ q^{\ 2} \ p^{\ 1} \ + \ q^{\ 1} \ \overline{p}^{\ 2} \ \right )
\vspace*{0.2cm} \\
\overline{o} \ = \ \left ( \ \overline{p} \ , \ - \ q \ \right )
\vspace*{0.2cm} \\
\rightarrow \ \mbox{for} \ o^{\ 2} \ = \ \overline{o}^{\ 1}
\hspace*{0.2cm} ; \hspace*{0.2cm} o^{\ 2} \ =
\ \left ( \ \overline{p}^{\ 1} \ , \ - \ q^{\ 1} \ \right )
\vspace*{0.3cm} \\
\begin{array}{lll}
o^{\ 1} \ \odot \ \left ( \ o^{\ 2} \ = \ \overline{o}^{\ 1} \ \right )
& \hspace*{-0.1cm} = \hspace*{-0.1cm}
& \left ( \ p^{\ 1} \ \overline{p}^{\ 1} \ + \ \overline{q}^{\ 1} \ q^{\ 1}
\ , \ - \ q^{\ 1} \ p^{\ 1} \ + \ q^{\ 1} \ \overline{\overline{p}}^{\ 1}
\ \right ) 
\vspace*{0.2cm} \\
& \hspace*{-0.1cm} = \hspace*{-0.1cm} & \left \lbrace 
\ \left | \ p^{\ 1} \ \right |^{\ 2} \ + \ \ \left | \ q^{\ 1} \ \right |^{\ 2}
\ \right \rbrace \ j_{\ 0} \ + \ 0
\end{array}
\vspace*{0.2cm} \\ \hline \vspace*{-0.2cm} \\
j_{\ 0} \ = \ \P \ , \ j_{\ 1} \ , \ j_{\ 7}
\hspace*{0.2cm} ; \ \hspace*{0.2cm}
j_{\ 1,2,3} \ \simeq \ i_{\ 1,2,3}
\end{array}
\end{equation}

\noindent
In eq. \ref{octeq:2} we used the involutory properties

\vspace*{-0.3cm}
\begin{equation}
\label{octeq:3}
\begin{array}{l}
\overline{\overline{q}} \ = \ q 
\hspace*{0.2cm} ; \hspace*{0.2cm} 
\overline{\overline{o}} \ = \ o
\end{array}
\end{equation}

\noindent
It follows that unitary quaternions 
( $q \ \overline{q} \ = \ \overline{q} \ q \ = \ \P$ )
are equivalent to $S_{\ 3} \ \simeq \ SU2 \ \subset \ R_{\ 4}$ , whereas
unitary octonions ( $o \ \overline{o} \ = \ \overline{o} \ o \ = \ \P$ ) are
equivalent to $S_{\ 7} \ \subset \ R_{\ 8}$ .
\vspace*{0.1cm} 

\noindent
This leads together with the complex numbers to the algebraic
association of $N \ = 1$ and $N \ = 2 \ -$ \hspace*{0.1cm}
$\Sigma$ variables to the
{\it three} inequivalent 'field'-algebras

\vspace*{-0.3cm}
\begin{equation}
\label{octeq:4}
\begin{array}{l}
\begin{array}{llll}
1 & N \ = \ 1 & \leftrightarrow & \mathbb{C} 
\ \simeq \ R_{\ 2} \ \supset \ S_{\ 1}
\vspace*{0.2cm} \\
2 & N \ = \ 2 & \leftrightarrow & \mathbb{Q} 
\ \simeq \ R_{\ 4} \ \supset \ S_{\ 3}
\vspace*{0.2cm} \\
3 & N \ = \ 2 & \leftrightarrow & \mathbb{O}
\ \simeq \ R_{\ 8} \ \supset \ S_{\ 7}
\end{array}
\end{array}
\end{equation}

\noindent
The group structures of cases 1 - 3 in eq. \ref{octeq:4} correspond to 

\vspace*{-0.3cm}
\begin{equation}
\label{octeq:5}
\begin{array}{lll llc}
1 & : & S_{\ 1} & \simeq \ U1 & \leftrightarrow & U1_{\ R} 
\ \otimes \ U1_{\ L} \\
2 & : & S_{\ 3} & \simeq \ SU2 & \leftrightarrow & SU2_{\ R} \ \otimes
\ SU2_{\ L} \\
3 & : & S_{\ 7} &            &  \leftrightarrow & U2_{\ L} 
\ \otimes \ U2_{\ R}
\end{array}
\end{equation}

\noindent
While the model introduced by M. Gell-Mann and M. L\'{e}vy \cite{LevyGM}
corresponds to case 2 ( eq. \ref{octeq:4} , \ref{octeq:5} ) , it is
case 3 ( also for $N \ = \ 2$ ) which is {\it different} and 
the {\it only} one extendable to $N \ > \ 2$ .

\noindent
This shall be illustrated for $N \ = \ 3$ and from there back to case 3 with 
$N \ = \ 2$ in the next section.

\newpage

\begin{center}
{\bf 1.3
$\Sigma \ = \ \frac{1}{\sqrt{2}} \ \left ( \ \sigma \ - 
\ i \ \pi \ \right )$ for $N \ = \ N_{\ fl} \ = \ 3$ 
( $m_{\ u} \ \sim \ m_{\ d} \ \sim \ m_{\ s}$ )}
\end{center}

\noindent
For $N \ = \ 3$ the $\Sigma \ -$ variables describe a $U3_{\ fl} \ -$ nonet of
{\it scalars and pseudoscalars} (one each) . I shall use the notation
$\Sigma_{\ \rightarrow \ \pi \ , \ K \ , \ \eta \ , \ \eta^{\ '}}$
labelled by the names of pseudoscalars , yet denoting associated pairs \\
{\it scalars $\leftrightarrow$ pseudoscalars}

\vspace*{-0.1cm}
\hspace*{-1.0cm}
\begin{equation}
\label{noneq:1c}
\begin{array}{l}
\Sigma \ =
\ \left (
\ \begin{array}{lll}
\Sigma_{\ 11} & \Sigma_{\ \pi^{\ -}} & \Sigma_{\ K^{\ -}}
\vspace*{1.0cm} \\
\Sigma_{\ \pi^{\ +}} & \Sigma_{\ 22} & \Sigma_{\ \overline{K}^{\ 0}}
\vspace*{1.0cm} \\
\Sigma_{\ K^{\ +}} & \Sigma_{\ K^{\ 0}} & \Sigma_{\ 33}
\end{array}
\right )
\vspace*{1.0cm} \\
\Sigma_{\ 11} \ = \ \frac{1}{\sqrt{3}} \ \Sigma_{\ \eta^{\ '}}
\ + \ \frac{1}{\sqrt{2}} \ \Sigma_{\ \pi^{\ 0}} \ + 
\ \frac{1}{\sqrt{6}} \ \Sigma_{\ \eta}
\vspace*{1.0cm} \\
\Sigma_{\ 22} \ = \ \frac{1}{\sqrt{3}} \ \Sigma_{\ \eta^{\ '}}
\ - \ \frac{1}{\sqrt{2}} \ \Sigma_{\ \pi^{\ 0}} \ +
\ \frac{1}{\sqrt{6}} \ \Sigma_{\ \eta}
\vspace*{1.0cm} \\
\Sigma_{\ 33} \ = \ \frac{1}{\sqrt{3}} \ \Sigma_{\ \eta^{\ '}}
\hspace*{2.3cm} - \ \frac{2}{\sqrt{6}} \ \Sigma_{\ \eta}
\end{array}
\end{equation}

\noindent 
In the chiral limit $m_{\ u,d,s} \ \rightarrow \ 0$ -- 8 pseudoscalar 
Goldstone modes become massless : $\pi \ , \ (3) \ ; 
\ K \ , \ \overline{K} \ , \ (4)
\ ; \ \eta \ , \ (1)$ , whereas $\eta^{\ '}$ and all 9 scalars 
remain massive.

\noindent
$\pi_{\ 0} \ \leftrightarrow \ \eta \ \leftrightarrow \ \eta^{\ '} \ -$
mixing -- 
eventually different for scalars relative to pseudoscalars --
is not discussed here \cite{WOPM} .
\vspace*{0.1cm} 

Projecting back on case 3 and $N \ = \ 2$ in the limit 
$m_{\ s} \ \rightarrow \ \infty$ an $SU2_{\ fl} \ -$ singlet {\it pair} 
-- denoted $\Sigma_{\ \eta_{\ (2)}}$ -- forms as ( singlet ) combinations 
of $\Sigma_{\ \eta} \ , \ \Sigma_{\ \eta^{\ '}}$ and a corresponding
isotriplet {\it pair} $\Sigma_{\ \pi} \ \rightarrow \ \vec{\Sigma}_{\ \pi}$ .
\vspace*{0.1cm}

\noindent
Instead of the $2 \ \times \ 2$ matrix form pertinent to 
case 3 and $N \ = \ 2$ we can equivalently display the {\it double
quaternion} basis from the octonion - structure ( eq. \ref{octeq:2} )

\vspace*{-0.1cm}
\begin{equation}
\label{noneq:2c}
\begin{array}{llll}
p & \leftrightarrow 
& \left ( \begin{array}{rcr} 
\sigma_{\ \eta_{\ (2)}} & , & \vec{\pi}
\end{array}
\hspace*{0.3cm} \right ) & \rightarrow \ \mbox{\cite{LevyGM}}
\vspace*{0.2cm} \\
q & \leftrightarrow 
& \left ( \begin{array}{rcr} 
\eta_{\ (2)} & \hspace*{0.3cm} , & \vec{\sigma}_{\ \pi} 
\end{array}
\right ) &
\end{array}
\end{equation}

\newpage

\begin{center}
\hspace*{-0.3cm} \section{\hspace*{-0.3cm}
From $\left \langle \ \Sigma \ \right \rangle$ as spontaneous real parameter
to $f_{\ \pi}$}
\label{three}
\end{center}

\noindent
As shown in section \ref{two} , the $\Sigma \ -$ variables are chosen such ,
that the spontaneous breaking of {\it just} chiral symmetry can be explicitely
realized .
\vspace*{0.1cm}

\noindent
For N equal ( positive ) quark masses it folows

\vspace*{-0.1cm}
\hspace*{-1.0cm}
\begin{equation}
\label{fpieq:1c}
\begin{array}{l}
\left \langle \ \Sigma \ \right \rangle \ = \ S \ \P_{\ N \times \ N}
\vspace*{1.0cm} \\
S \ = \ \frac{1}{\sqrt{2 \ N}} 
\ \left \langle \ \sigma^{\ 0} \ \right \rangle
\hspace*{0.5cm} ; \hspace*{0.5cm} \Sigma \ = \ \frac{1}{\sqrt{2}}
\ \left ( \ \sigma \ - \ i \ \pi \ \right )_{\ N \times N}
\vspace*{1.0cm} \\
\begin{array}{lll}
j_{\ \mu \ R}^{\ a} & = & i \ S \ tr \ \frac{1}{2} 
\ \lambda^{\ a} \ \partial_{\ \mu}
\ \left ( \ \Sigma \ - \ \Sigma^{\ \dagger} \ \right )
\hspace*{0.2cm} + \ \cdots
\vspace*{1.0cm} \\
& = & S \ \partial_{\ \mu} \ \pi^{\ a} 
\hspace*{0.2cm} + \ \cdots
\end{array}
\vspace*{1.0cm} \\
\Sigma \ - \ \Sigma^{\ \dagger} \ = \ - \ i \ \pi^{\ b} \ \lambda^{\ b}
\vspace*{1.0cm}  \\
\left \langle \ \Omega \ \right | \ j_{\ \mu \ R}^{\ a} 
\ \left | \ \pi^{\ b} \ , \ p \ \right \rangle \ = \ i \ \frac{1}{2}  
\ f_{\ \pi} \ p_{\ \mu} \ \delta^{\ ab} 
\hspace*{0.2cm} \mbox{for} \hspace*{0.2cm} a,b > 0
\vspace*{1.0cm}  \\
\ - \ S \ = \ \frac{1}{2} \ f_{\ \pi}
\hspace*{0.2cm} \leftrightarrow \hspace*{0.2cm}
- \ \left \langle \ \sigma^{\ 0} \ \right \rangle \ = 
\ \left ( \ \frac{N}{2} \ \right )^{\ 1/2} \ f_{\ \pi}
\hspace*{0.2cm} ; \hspace*{0.2cm}
f_{\ \pi} \ \sim \ 92.4 \ \mbox{\begin{tabular}[t]{c}MeV \vspace*{0.1cm} \\
for $\vec{\pi}$ \end{tabular}
}
\end{array}
\end{equation}

\begin{center}
{\bf Acknowledgement
}
\end{center}

\noindent
The present account was the subject of a lunch-seminar, 16. April 2008 in Bern.
The discussions especially with Uwe-Jens Wiese, \'{E}milie Passemar and
Heinrich Leutwyler are gratefully acknowledged .

\end{document}